\newcommand{\CLASSINPUTtoptextmargin}{.75in}
\newcommand{\CLASSINPUTbottomtextmargin}{.98in}
\documentclass[conference]{IEEEtran}
\IEEEoverridecommandlockouts
\ifCLASSINFOpdf
\else
\fi

\usepackage[caption=false]{subfig}
\usepackage{cite}
\usepackage{upgreek}
\usepackage[pdftex]{graphicx}
\usepackage{cite}
\usepackage{xcolor,soul,framed} 
\colorlet{shadecolor}{yellow}
\usepackage{url}
\usepackage{array}
\usepackage{caption}
\usepackage{balance}
\usepackage{cite}
\usepackage{xcolor,soul,framed} 
\colorlet{shadecolor}{yellow}
\usepackage{color,soul}
\usepackage[pdftex]{graphicx}
\DeclareGraphicsExtensions{.pdf,.jpeg,.png}
\usepackage[cmex10]{amsmath}
\usepackage{algorithmic}
\usepackage{mdwmath}
\usepackage{mdwtab}
\usepackage{eqparbox}
\usepackage{url}
\usepackage{array}
\usepackage{fixltx2e}
\usepackage{dblfloatfix}
\usepackage[mathscr]{euscript}
\usepackage{amsfonts}
\usepackage{amsmath}
\usepackage{caption}
\usepackage{commath}
\usepackage{upgreek}
\usepackage[long]{optidef}
\usepackage{bm}
\usepackage{algorithm}

\usepackage[open,openlevel=2]{bookmark}
\usepackage{balance}
\usepackage{textgreek}
\usepackage{booktabs,chemformula}
\usepackage{array}
\usepackage{float}
\usepackage{amssymb}
\usepackage{graphicx}
\usepackage{bm}
\usepackage{cleveref}
\usepackage[thinlines]{easytable}
\usepackage{gensymb}
\usepackage{lipsum}  
\usepackage{comment}

\makeatletter
\def\footnoterule{\kern-3\p@
  \hrule \@width 2in \kern 2.6\p@} 
\makeatother
\begin{document}
\IEEEoverridecommandlockouts

\title{Toward cm-Level Accuracy: Carrier Phase Positioning for IIoT in 5G-Advanced NR Networks}
\vspace{-5 in}
\author{\IEEEauthorblockN{Abdurrahman Fouda, Ryan Keating,
and Hyun-Su Cha 
}
\IEEEauthorblockA{Nokia Standards\\
abdurrahman.fouda@u.northwestern.edu, ryan.keating@nokia.com, hyun-su.cha@nokia.com
}}

\maketitle
\begin{abstract}
High-precision positioning accuracy is among the key features of the future fifth-generation (5G-advanced) cellular networks to enable a wide variety of commercial, critical, and consumer use cases. 5G new radio (NR) systems have relied on (1) cellular temporal/angular-based positioning methods to provide the indoor environments with a moderate positioning accuracy that is well below the positioning requirements of these future use cases and (2) highly precise satellite carrier phase/code-based positioning methods for the outdoor deployments that are limited by the availability of the satellite coverage. This paper defines the relevant standard mechanisms and algorithms to use the carrier phase cellular-based measurements as a potential solution to achieve a high-precision positioning estimation accuracy in 5G-advanced NR networks. The presented positioning technique is evaluated using high-fidelity system-level simulations for indoor factory (InF) deployment scenarios. The numerical results demonstrate that the presented technique can significantly improve the positioning accuracy compared with the state-of-the-art NR positioning methods. Our findings in this paper also show that the carrier phase method not only provides an indoor complement to the outdoor satellite positioning but also provides an outdoor alternative to the high-precision satellite methods.
\end{abstract}

\IEEEpeerreviewmaketitle

\section{Introduction}\label{sec_intro}
The ability to achieve stable and high-precision positioning accuracy continues to become more important for different verticals and services in the fifth generation (5G) new radio (NR) cellular networks. These applications include but not limited to commercial use cases (e.g., industrial internet of things (IIoT), full range of asset tracking and digital twins), critical use cases (e.g., vehicular networks), and consumer use cases (e.g., virtual/augmented reality gaming). In this regard, the 3rd Generation Partnership Project (3GPP) has introduced, in its release 16 (Rel-16), the minimum performance targets for NR positioning in indoor and outdoor use cases that range from 10 to 3 m (80\% availability) for horizontal and vertical positioning errors, respectively~\cite{38855}. Depending on the service level (i.e., the nature of the operation) of each use case, the accuracy requirements for horizontal and vertical positioning are expected to continue to become stricter in the 5G-advanced networks (known as Rel-18) and beyond. For example, the horizontal and vertical accuracies in service level 6 use cases vary from .3 to 2 m (99.9\% availability), respectively~\cite{22261}.    

Over the last two decades, users have relied on the global navigation satellite system (GNSS) to achieve high positioning accuracy in outdoor use cases. Currently, the state-of-the-art in positioning techniques for outdoor applications is real-time kinematic (RTK) GNSS~\cite{ESA}. RTK-GNSS uses the so-called \textit{carrier phase measurement} method to achieve up to 10 cm positioning accuracy. Specifically, the receiver uses the collected phase measurements from different satellites along with their known locations for positioning estimation. However, RTK-GNSS methods suffer from high deployment costs and only work in outdoor environments with good line of sight (LOS) connections to multiple satellites. Hence, it becomes challenging to achieve high accuracy using the GNSS-based positioning methods in weak LOS (NLOS) scenarios like dense urban areas, indoor environments, tunnels, and underground parking (see~\cite[Sec.~7.2]{38855} and references therein).     
 
Recently, cellular-based (i.e., RAT-dependent) positioning technologies have emerged as potential solutions to complement the GNSS-based technologies in the areas where GNSS coverage is not available. These positioning solutions (which have been specified for NR Rel-16 and beyond) utilize the timing measurements to locate the user equipment (UE) and are categorized into temporal methods like the downlink/uplink time difference of arrival (DL/UL-TDOA) and multi-cell round trip time (Multi-RTT) methods, angular methods like the downlink/uplink angle of departure/arrival (DL/UL-AoD/AoA) methods, and hybrid schemes~\cite{Ryan2}. Furthermore, there have been many contributions in the state-of-the-art literature for hybrid NR and GNSS positioning to improve the positioning accuracy in NR networks~\cite{SPNTV}. However, none of these methods support the required cm-level precision for highly-accurate service levels (e.g., IIoT) based on the evaluation results reported in~\cite{qcom,Nokia}. In addition, some of these RAT-dependent positioning methods can only achieve their best performance assuming perfect synchronization between base stations (gNBs) which is difficult to be achieved in real network deployments. In an ideal case, the availability of high-precision NR positioning should be independent of the environment (e.g., indoor or outdoor) and resilient to the loss of other technologies (i.e., GNSS). 

In this paper, we investigate the use of carrier phase-based positioning as a potential solution to improve the positioning estimation accuracy in 5G-advanced NR networks. Specifically, we propose the standard-relevant mechanisms and algorithms required to implement carrier phase measurements in gNBs and UEs in NR networks. Furthermore, we rigorously evaluate the overall system performance against the Rel-16/17 baseline using high-fidelity system simulations and the state-of-the-art absolute time of arrival (ToA) 3GPP channel model for indoor factory sparse high (InF-SH) deployment scenarios~\cite{38901}. The simulation results demonstrate that the carrier phase method can significantly improve the positioning estimation accuracy (by a factor of $\approx$ 10) compared with Rel-16/17 methods. Our analysis in this paper show that carrier phase-based positioning enables the 5G-advanced NR networks to offer high-precision positioning using only a single wireless system (i.e., the cellular one) and be resilient to the loss of satellite-based positioning. The use of carrier phase method for NR positioning has been briefly discussed in~\cite{CATT}. A carrier phase-based ToA estimation has been presented in~\cite{ranging} for NR networks. To the best of our knowledge, none of the prior art has neither identified the related standard mechanisms to enable carrier phase positioning in 5G-advanced networks nor evaluated their overall system-level performance using the state-of-the-art industrial 3GPP channel model~\cite{38901}.

\section{NR Downlink Carrier Phase Positioning}\label{sec_cpmethod}
In this section, we discuss the relevant algorithms for carrier phase positioning and show how they can be applied in NR networks. NR Carrier phase positioning is expected to involve the transmission of carrier phase positioning reference signals (CP-PRS). CP-PRS may be the same as the existing Rel-16/17 PRS for positioning or it may be redesigned specifically for the NR carrier phase technique. CP-PRS can be a pure carrier wave of periodic wide-band sinusoidal signals or continuous narrow-band signals transmitted at a pre-defined carrier frequency~\cite{CATT}. For simplicity, we will call the reference signal CP-PRS in this paper which is used to collect the carrier phase measurements for positioning estimation. 

\subsection{Carrier Phase Concept }\label{subsec_concept}
Carrier phase-based positioning relies on the idea of mixing the reference signal (generated at the transmitter) with its replica at the receiver to generate a mixed signal with low and high-frequency components. The high-frequency component can be filtered-out (at the receiver), leaving only a carrier signal whose phase is the difference between the phase of the transmitted signal and its replica at the receiver~\cite{ESA}. In ideal settings, the relation between the phase difference (i.e., $\varphi$) and the geometric distance between transmitter and receiver (i.e., $d$) is determined by $\varphi=2{\pi}d/\lambda$, where $\lambda$ represents the wavelength of the operating carrier frequency. In NR networks, the phase difference can be used to estimate the distance between the $i^{\text{th}}$ transmitter and $m^{\text{th}}$ receiver as follows:
\begin{equation}\label{eq1}
\frac{\lambda}{2\pi}{\varphi_{}}_{i,m}= {d_{}}_{i,m}+c\left({b_{}}_{m}-{b_{}}_i\right)+\lambda{{N_{}}_{i,m}}+{\nu_{}}_{i,m},
\end{equation}
where we denote by $c$ the speed of light, ${b_{}}_i$ the $i^{\text{th}}$ transmitter clock bias and ${b_{}}_m$ the $m^{\text{th}}$ receiver clock bias. ${d_{}}_{i,m}$ and ${\nu_{}}_{i,m}$ represent the geometric distance, and phase measurement errors between the $i^{\text{th}}$ transmitter and $m^{\text{th}}$ receiver, respectively. For convenience, we replace the \textit{transmitter} term with gNB and the \textit{receiver} term with UE. Note that the left-hand side of~(\ref{eq1}) is in meters. Here, ${N_{}}_{i,m}$ represents the unknown integer ambiguity parameter which is the total number of complete phase cycles that the reference carrier signal has travelled between UE and gNB to produce the same observed phase at the UE. Essentially, the phase measurements are ambiguous because we measure the amplitude of a periodic signal. It is quite challenging to estimate this number given that the phase repeats itself every complete cycle (i.e., $2\pi$). 

There have been several approaches to solve the integer ambiguity problem for carrier phase positioning in the GNSS networks. These approaches include the well-known least-squares ambiguity decorrelation adjustment (LAMBDA) method which can help to maintain the probability of wrong fixing for the integer ambiguity well below $10^{-6}$ (see~\cite{IAR} and references therein). Recently, the use of the virtual phase measurements with the virtual wavelength has been proposed in~\cite{CATT} to resolve the integer ambiguity problem in 5G NR networks. Note that this method means that the time domain measurements are essentially used to resolve the integer ambiguity. However, this method requires the transmission of the CP-PRS signals over two or more frequencies which may not always be feasible from a spectrum efficient utilization perspective. Since the LAMBDA method can provide a very good performance and given that the time domain measurements can be used to resolve the integer ambiguity problem (e.g., in case of the virtual wavelength method), our focus in this paper is to initially evaluate the carrier phase method with an ideal integer ambiguity resolution. Hence, the estimated phase of CP-PRS signal at $i^{\text{th}}$ UE can be written as:
\begin{equation}\label{eq2}
\frac{\lambda}{2\pi}\tilde{\varphi_{}}_{i,m}= {d_{}}_{i,m}+c\left({b_{}}_m-{b_{}}_i\right)+{\nu_{}}_{i,m}.
\end{equation}


\subsection{Differencing Operations}\label{subsec_diffops}
As shown in~(\ref{eq2}) the unknown phase shifts due to the oscillator phases at the gNB and UE can add additional ambiguity to the estimated phase at UE and impair the estimation accuracy of the phase-based positioning. Unknown frequency offsets (i.e., time-dependent phase errors) can be kept small by synchronization or compensated by adding a constant phase error to the frequency phase offset measurements. Intuitively, it is very difficult to achieve perfect synchronization between all network elements. Further, the phase offset compensation does not give the most accurate estimation for unknown number of complete travelled phase cycles. Hence, the use of the so-called \textit{differencing operations} is considered as a potential candidate to mitigate the impacts of unknown phase shifts during the phase estimations. 

\begin{figure}
  \begin{center}
  \includegraphics[width=8cm,height=8cm,,keepaspectratio]{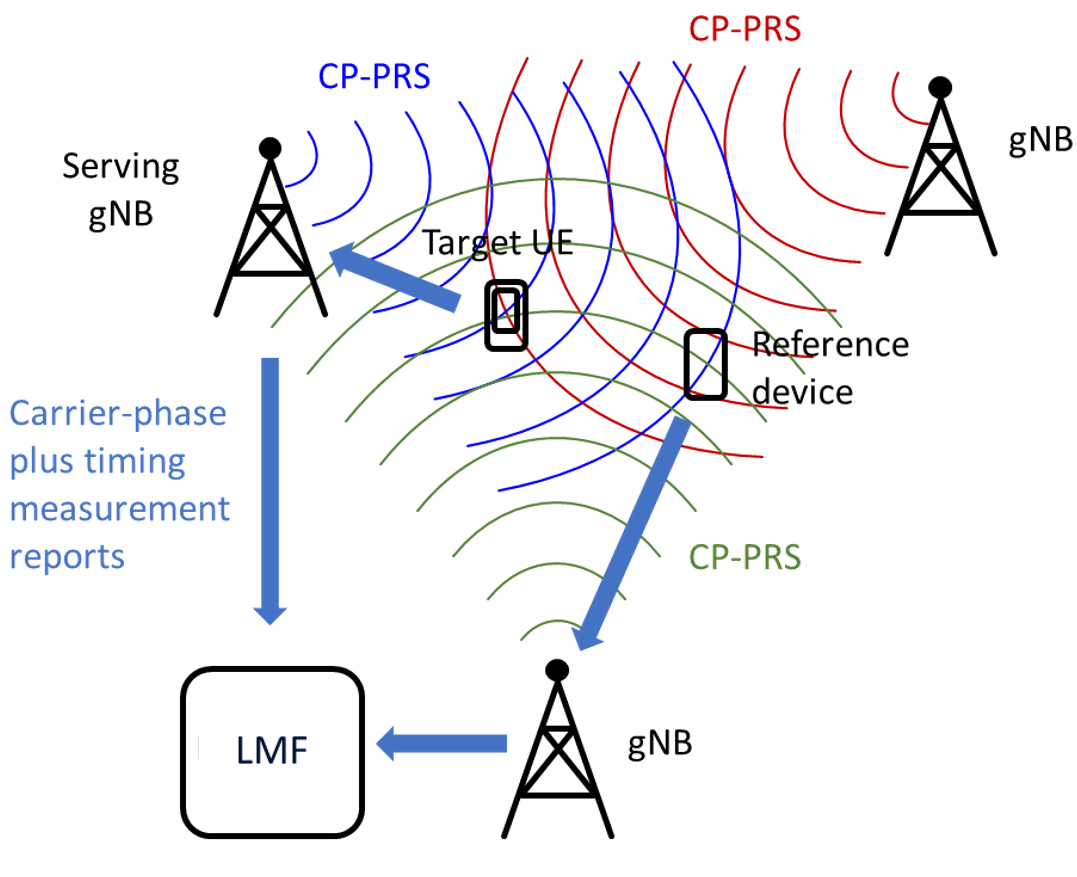}
  \caption{Expected NR setup for carrier phase positioning.}\label{fig_illust}
  \vspace{-.25in}
  \end{center}
\end{figure}

The concept of differencing operations relies on the idea of differencing the phase measurements of the target receiver and a reference node (i.e., network element) with a known fixed location to eliminate the unknown clock bias errors. Differencing operations involve two operations: namely, \textit{single differencing} and \textit{double differencing}. Single differencing operation removes the clock bias errors between UE and gNB and double differencing removes the clock bias errors between different gNBs. In this paper, we use the term \textit{target receiver} to refer to UE for which the position is being estimated. The differenced measurements are then used to estimate the relative UE position with respect to the reference node’s location. Given that the reference node’s location is fixed and known, the exact position of the target UE can be easily calculated using any trilateral estimation algorithm. It is expected that the reference node will also measure the carrier phase on the CP-PRS. In this paper, we propose that the reference node can be another UE and refer to it by \textit{reference UE}. Fig.~\ref{fig_illust} shows the expected reference signal transmission between reference, target UEs, serving, and neighbor gNBs. It also shows the transmission of the measurement reports to the location server (LMF). Note that the signal processing operations in this setup are distributed between different network elements (i.e., not centralized in a single network entity).


Now, let $\tilde{\varphi_{}}_{i,m}$ and $\tilde{\varphi_{}}_{j,m}$ (defined by means of~(\ref{eq2})) denote the phase measurements of the $m^{\text{th}}$ target UE from the $i^{\text{th}}$ and $j^{\text{th}}$ gNBs, respectively where $j^{\text{th}}$ gNB is the serving gNB for that UE. The single differenced phase is given by:

\begin{equation}\label{eq3}
\frac{\lambda}{2\pi}\Delta\tilde{\varphi}_{m}= \Delta{d_{}}_{m}+c{b_{}}_{j,i}+\Delta{\nu_{}}_{m},
\end{equation}
where $\Delta\tilde{\varphi_{}}_{m}=\tilde{\varphi_{}}_{i,m}-\tilde{\varphi_{}}_{j,m}$, $\Delta{{d_{}}_{m}}={{d_{}}_{i,m}}-{{d_{}}_{j,m}}$, ${b_{}}_{j,i}={b_{}}_j-{b_{}}_i$ and $\Delta{\nu_{}}_{m}={\nu_{}}_{i,m}-{\nu_{}}_{j,m}$. Note that the ${b_{}}_{m}$ clock bias of the $m^{\text{th}}$ UE is removed using~(\ref{eq3}). We propose that the serving gNB can be used as a reference transmitter for all its serving UEs (UEs that are in its serving sector) and use the phase measurements with it to generate their single differencing measurements from all other detected neighbor gNBs. Similarly, the single differenced phase measurements of the $n^{\text{th}}$ reference UE can be denoted by $\Delta\tilde{\varphi_{}}_n$ where $\Delta\tilde{\varphi_{}}_n=\tilde{\varphi_{}}_{i,n}-\tilde{\varphi_{}}_{j,n}$. The double differenced measurements can be generated then by adding another step of differencing between the single-differenced measurements of the of the target $m^{\text{th}}$ UE and $n^{\text{th}}$ reference UE as follows:
\begin{equation}\label{eq4}
\frac{\lambda}{2\pi}\nabla\Delta\tilde{\varphi}= \nabla\Delta{d}+\nabla\Delta\nu,
\end{equation}
where $\nabla\Delta\tilde{\varphi}=\Delta\tilde{\varphi_{}}_{m}-\Delta\tilde{\varphi_{}}_{n}$, $\nabla\Delta{d}=\Delta{d_{}}_{m}-\Delta{d_{}}_{n}$ and $\nabla\Delta\nu=\Delta{\nu_{}}_{m}-\Delta{\nu_{}}_{n}$. As can be noted in~(\ref{eq4}), the fractional parts of the unknown oscillator phase shifts (i.e., ${b_{}}_{i}$ and ${b_{}}_{j}$) are eliminated using a reference UE with a known fixed position and double differencing operations. By means of~(\ref{eq4}), the differential distance of the $m^{\text{th}}$ target UE can be given by:

\begin{equation}\label{eq5}
\Delta{d_{}}_{m}=\frac{\lambda}{2\pi}\nabla\Delta\tilde{\varphi}+\Delta{d_{}}_{n}-\nabla\Delta\nu.
\end{equation}

At least two additional similarly refined phase measurements are collected at the target UE or LMF to form a system of linearly independent equations which can be solved using the least-squares (LS) method as will be discussed in the next Section. We use the term \textit{refined phase measurements} to denote the phase measurements without frequency offset impairments and integer ambiguity parameters (e.g., $\nabla\Delta\tilde{\varphi}$).     

\section{LS Method for Positioning Estimation using Carrier Phase Double-differenced Measurements}\label{sec_LS}
We propose that the Taylor series expansion method can be used to solve the positioning estimation problem using the LS method as described in~\cite{LSEst}. We start with the serving gNB position $\mathbf{d_{}}_{\text{s}}=\left[{x_{}}_{\text{s}},{y_{}}_{\text{s}},{z_{}}_{\text{s}}\right]$ as an initial position guess of the target UE, and update the position deviation at the $k^{\text{th}}$ iteration of the least-squares algorithm as follows: 

\begin{equation}\label{eq6}
    \Delta\mathbf{d_{}}_{\text{s}}^{(k)}=\left(\mathbf{G}^{\top}\mathbf{G}\right)^{-1}\mathbf{G}^{\top}\mathbf{h}^{\top},
\end{equation}
where $\mathbf{h}\in\mathbb{R}^{1\times{I}}$ denotes the vector containing the double-differenced residual observations and is given by $\mathbf{h}=\left({h_{}}_{m,i}:i\in\mathscr{J}\right)$. Note that $\mathscr{J}$ represents the set of detected phase measurements by $m^{\text{th}}$ target UE with cardinality $|\mathscr{I}|$ where $|\mathscr{I}|=I$. The values ${h_{}}_{m,i}$ can be calculated as: 

\begin{equation}\label{eq7}
    {h_{}}_{m,i}=\frac{\lambda}{2\pi}\nabla\Delta{\varphi_{}}_{\text{r}}+\lambda{N_{}}_{\text{r}},
\end{equation}
where $\nabla\Delta{\varphi_{}}_{\text{r}}$ is the differential residual phase measurement between $m^{\text{th}}$ UE and $i^{\text{th}}$ gNB which is given by $\nabla\Delta{\varphi_{}}_{\text{r}}=\nabla\Delta{\varphi_{}}_{\text{e}}-\nabla\Delta{\varphi_{}}_{{k}}$ where $\nabla\Delta{\varphi_{}}_{\text{e}}$ is the estimated double differenced phase measurement and $\nabla\Delta{\varphi_{}}_{{k}}$ is the computed double differenced phase measurement at the $k^{\text{th}}$ iteration of the least-squares algorithm. We denote by $\mathbf{G}\in\mathbb{R}^{I\times{3}}$ the design matrix where $\mathbf{G}=[{\mathbf{g_{}}}_{x},\,{\mathbf{g_{}}}_{y},\,{\mathbf{g_{}}}_{z}]$ with ${\mathbf{g}_{}}_{x}$ representing the design vector of the x-coordinate. In particular, ${\mathbf{g}_{}}_{x}$ can be given by ${\mathbf{g}_{}}_{x}=\left({g_{}}_{x,i}:i\in\mathscr{I}\right)$ where we calculate ${g_{}}_{x,i}$ as follows:
\begin{equation}\label{eq8}
    {g_{}}_{x,i}=\frac{{x_{}}_m-{x_{}}_i}{{d_{}}_{i,m}}-\frac{{x_{}}_m-{x_{}}_s}{{d_{}}_{s,m}},
\end{equation}
where ${x_{}}_m$, ${x_{}}_i$ and ${x_{}}_{\text{s}}$ are the x-coordinates of the $m^{\text{th}}$ target UE, $i^{\text{th}}$ gNB and the serving gNB (of the $m^{\text{th}}$ UE), respectively. We denote by ${d_{}}_{i,m}$ the geometric distance between $m^{\text{th}}$ UE and $i^{\text{th}}$ gNB. Similarly, ${d_{}}_{\text{s},m}$ denote the geometric distance between $m^{\text{th}}$ UE and its serving gNB. The least-square iterations are executed until the $l^2$-norm of the $\Delta{d_{}}_{s}^{(k)}$ is less than a small error deviation $\epsilon$.

It is worth mentioning that the discussed positioning algorithm for carrier phase method in NR networks can be implemented at either the target UE or the LMF based on the positioning mode. Generally, there are two positioning modes for 5G NR networks namely: a) UE-assisted mode, and b) UE-based mode. In the UE-assisted mode, UE uses CP-PRS to provide the carrier phase measurements from the serving and neighboring gNBs to the LMF. LMF then uses a trilateral estimation algorithm to estimate the position of the UE given the known locations of the gNBs, reference UE and the double differenced carrier phase measurements. In the UE-based mode, the UE is assumed to be capable of performing the carrier phase-based positioning estimation locally~\cite{Ryan2}. 

\section{System-level Evaluations}\label{sec_SLS}

Now, we evaluate the achievable gains of the positioning estimation accuracy using the carrier phase measurements that may be introduced in 5G-advanced NR networks (Rel-18 and beyond). Specifically, we use a high-fidelity system-level simulator that closely follows the 3GPP Rel-17 specified physical and medium access control (MAC) procedures~\cite{38201,38321}.
 The numerical results are generated for the IIoT use case of the InF-SH scenario. First, we analyze the performance of the carrier phase measurements using the wideband CP-PRS at an operating frequency of 3.5 GHz with a bandwidth of 100 MHz. We then evaluate the positioning estimation accuracy using the carrier phase measurements with the assumption of an ideal resolution for the integer ambiguity problem. In addition, we investigate the sensitivity of the positioning accuracy (at the horizontal and vertical levels) for the integer ambiguity errors. Finally, we comment on how the carrier phase method can achieve a high-precision accuracy compared with the reported accuracy of the current time-based positioning methods in~\cite{Nokia} using the same simulation assumptions. 

\begin{figure}
  \begin{center}
  \includegraphics[width=7cm,height=7cm,,keepaspectratio]{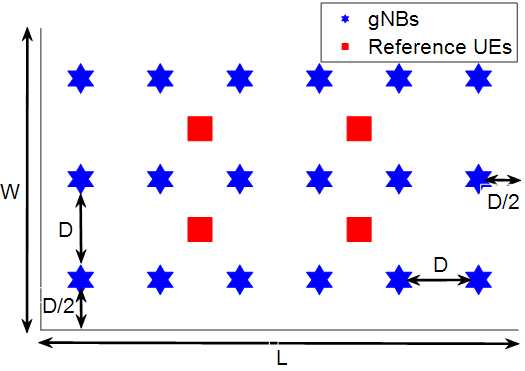}
  \caption{Factory 2D layout.}\label{fig_layout}
  \vspace{-.15in}
  \end{center}
\end{figure}

UEs use CP-PRS to collect the phase measurements on top of the time measurements from the serving and neighbor gNBs that are distributed over the factory hall using the absolute ToA 3GPP channel model~\cite{38901,5GCMIIoT}. We use a modified InF-SH scenario, in which, four reference UEs are integrated as new additional network elements and distributed equally over the factory hall to enable the double-differencing operations as shown in Fig.~\ref{fig_layout}. In other words, we do not assume perfect synchronization between the network elements. Note that the serving gNB is used as a reference transmitter to perform the single differencing operations (i.e., there is no need to deploy an additional network element as a reference transmitter). gNB heights are chosen randomly at uniform between $\left[\alpha,\,\beta\right]$ where $\alpha$ and $\beta$ are fixed integers with $0<\alpha<\beta$. Each gNB is equipped with a two-element cross-polarized antenna array.

We assume that the measured phase values depend on only the distance between gNB and UE (i.e., any phase impact from multiple antennas is compensated at gNB). The phase measurements of the target and reference UEs are then collected by either the target UE or LMF (see Section~\ref{sec_cpmethod}) to perform the related positioning operations. These operations include the single/double differencing and the resolution of the carrier ambiguity problem. Given the known locations of gNBs and the set of refined double-differenced measurements, the LS method is used to solve the positioning estimation problem. Results are collected from only UEs that fall in the convex hull of the factory to eliminate the edge effect.

\begin{table}
\renewcommand{\arraystretch}{1.3}
  \centering
  \caption{Simulation parameters.}\label{tab_simpara}
        \begin{tabular}{l l}
        \hline
            \textbf{Parameter}  & \textbf{Value}\\
        \hline
            Layout parameters & Big hall: L=300m$\times$W=150M, $\alpha=$3m,\\
            &$\beta=$10m, D=50m, Ceiling height=10m,\\
            &gNBs:18, Penetration loss: 0 dB\\&Floors: 1 ~\cite{Nokia}.\\
        \hline
            Channel model  & 3GPP, absolute ToA model~\cite{38901}, Cross-\\&correlation peak detection w/ threshold.\\
        \hline
           Carrier settings   & FR1, BW: 100MHz at 3.5GHz, NR\\
           &numerology: 30KHz subcarrier spacing.\\
        \hline
            Reference signal  & NR PRS, comb-6, Power-boost: 0dB.\\
        \hline
            gNB configurations   & Tx power: 24 dBm, Antenna geometry:\\
            &2-element XP [0,-45, 0, 45], Sectors: 1.\\
        \hline
            UE configurations  & Height: 1.5m, Walking, 1-Tx antenna.\\
        \hline
            Positioning methods  & Carrier-phase, DL-TDOA, 
            LS~\cite{LSEst}.\\
        \hline
        %
        \end{tabular}
\end{table}

\subsection{Performance of Phase Measurements}\label{subsec_cpmeas}
In this section, we evaluate the error rate of the collected phase measurements that are used for carrier phase-based positioning. We collect the measurements for two propagation InF-SH scenarios: a) LOS scenario, in which, the communication link between gNB and UE always has LOS connection (even if its signal strength is small), b) LOS/NLOS scenario, in which, UEs do not have LOS connections with some gNBs. To obtain the carrier phase measurement for gNBs, we use the cross-correlation peak detection method. In our simulations, UE identifies the LOS path based on a certain threshold. If UE detects a signal path power which is over a certain ratio of the maximum peak power among all signal paths, UE determines it as a LOS path. By using the cross-correlation of the received signals and the known PRS sequence, the carrier phase component is extracted from the peak of the cross-correlation coefficient of the identified LOS tap to calculate the phase measurement value. In that, we assume that UE knows the overall information that affects the phase of the LOS channel component ~\cite[Sec.~7.6.9]{38901}. We then calculate the phase measurement error by finding the absolute difference between the true and measured phase values. The true phase value is given by ${\varphi_{}}_{\text{t}}=2\pi{{d_{}}_{\text{t}}}/\lambda$, where ${d_{}}_{\text{t}}$ denotes the actual geometric distance between UE and gNB. 

\begin{figure}
  \begin{center}
  \includegraphics[width=7cm,height=7cm,,keepaspectratio]{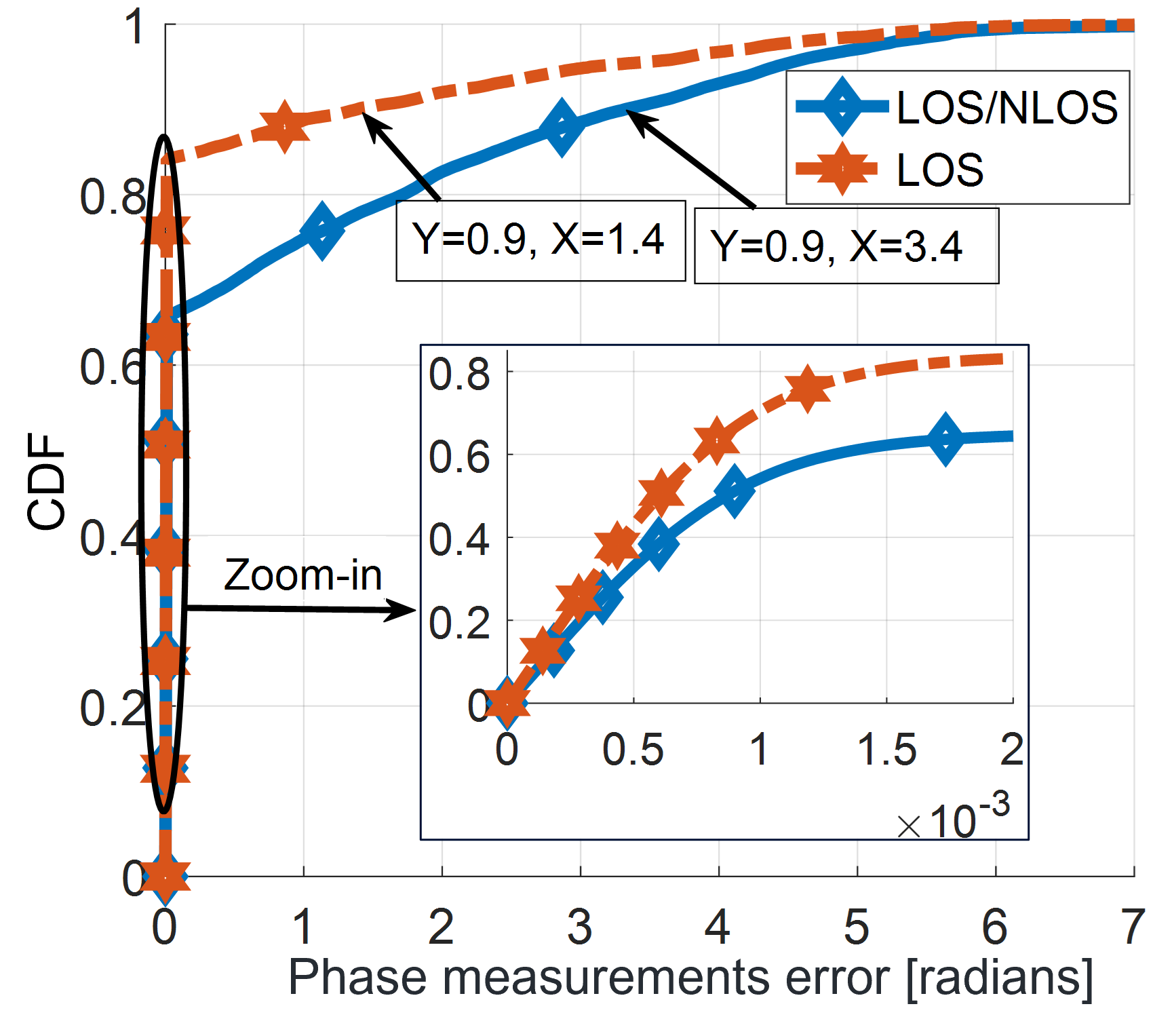}
  \caption{Phase measurements error: LOS vs. LOS/NLOS.}\label{fig_cpmeas}
  \vspace{-.25in}
  \end{center}
\end{figure}

Fig.~\ref{fig_cpmeas} demonstrates that the phase error is almost 1.4$\degree$ and 3.4$\degree$ in LOS and LOS/NLOS propagation environments, respectively. These angles correspond to $\approx$ 1.9 and 4.6 cm in distance, respectively. In particular, the phase error $\left({\delta_{}}_{\text{e}}\right)$ can be mapped into distance (${d_{}}_{\text{e}}$) using the following formula ${d_{}}_{\text{e}}=\lambda{\delta_{}}_{\text{e}}/2\pi$ under the assumption of a perfect resolution for the integer ambiguity problem. It is worth mentioning that Fig.~\ref{fig_cpmeas} depicts the phase error of the double differential carrier phase measurements (which are used by the positioning estimation algorithm), not the direct phase measurements. This is to get the most accurate estimation for the expected measurement error in the InF-SH deployment scenarios. Note also that, as expected, the double differential operation increases the observed measurement phase error. Our system simulations show that a phase measurement error of $\approx$ 2.3$\degree$ can be achieved for 90\% of UEs when the direct phase measurements are used compared with an error of 3.4$\degree$ with the double differential measurements. Fig.~\ref{fig_cpmeas} reveals that the carrier phase method can significantly improve the measurement accuracy of the positioning measurements compared with the conventional 3GPP time-based positioning methods~\cite{Nokia}. As shown, a sub-5 cm phase measurement error can be achieved for 90\% of the phase measurements making it possible to achieve high-precision accuracy using the carrier phase method. 

\subsection{Carrier Phase Positioning Accuracy Performance}\label{subsec_cpperf}
\begin{figure}
  \begin{center}
  \includegraphics[width=8.5cm,height=8.5cm,,keepaspectratio]{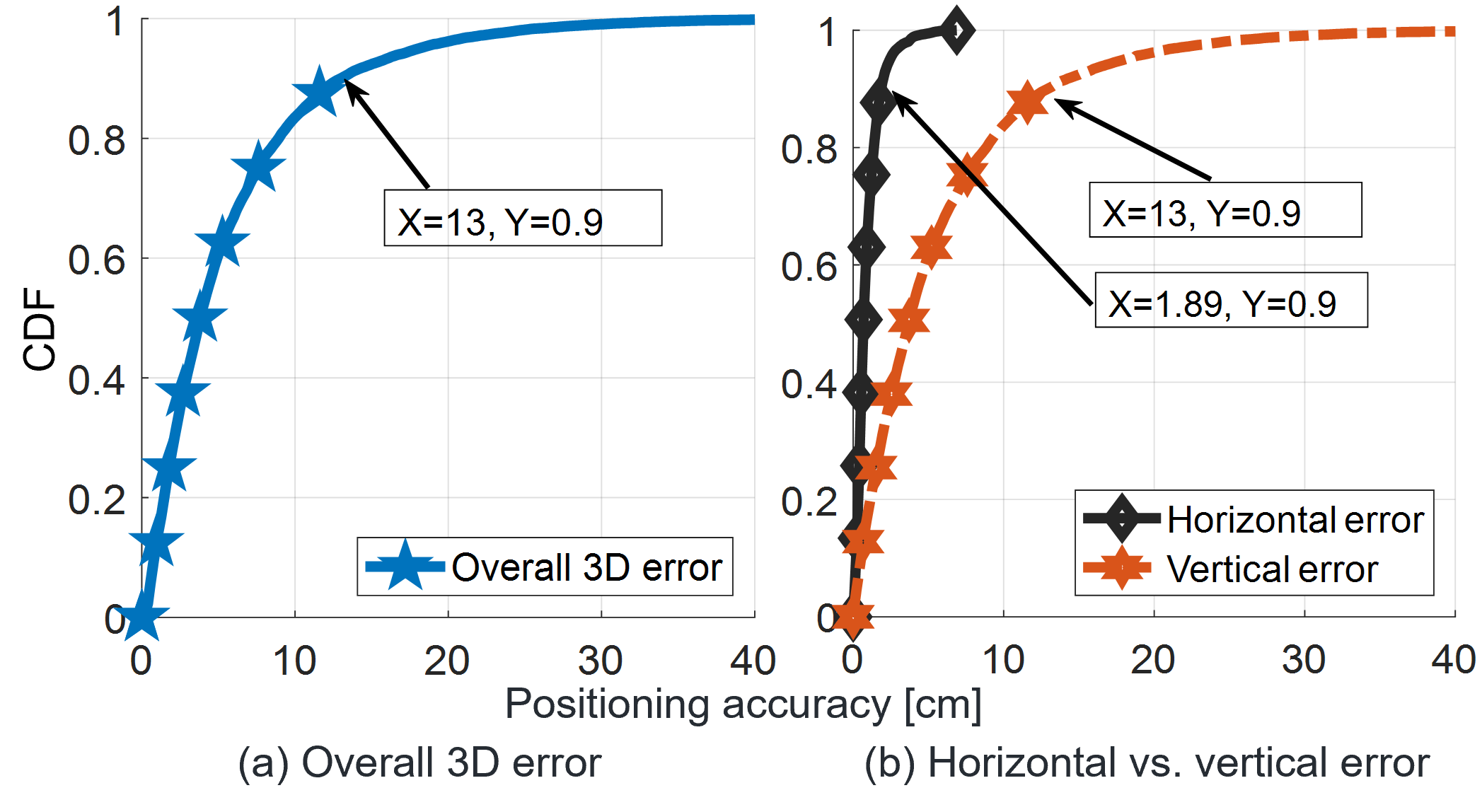}
  \caption{Carrier phase positioning accuracy.}\label{fig_cpacc}
  \vspace{-.25in}
  \end{center}
\end{figure}

In this section, we show that a high-precision positioning estimation accuracy (well beyond the Rel-16/17 requirements) can be achieved using the carrier phase method. As shown in Fig.~\ref{fig_cpacc}-a, the carrier-phase method can achieve a sub-15 cm and sub-10 cm overall 3D positioning accuracy for 90\% and 80\% of UEs, respectively. This confirms that the carrier phase positioning method can significantly improve the positioning estimation accuracy in NR networks compared with the current 3GPP angular and time positioning methods. Note that the results in this section are generated for LOS/NLOS environments and with the assumption of an ideal resolution to the integer ambiguity problem. In this paper, we use the term overall 3D position to denote the estimated $\left[x,y,z\right]$ coordinates of the UE. Our findings in this section show that the carrier phase method can provide highly accurate 3D positioning which is important for many use cases in the current/future 3GPP releases (e.g., IIoT, full range of asset tracking, and digital twins). 

We dig deeper into the accuracy of the estimated overall 3D position and focus on its horizontal and vertical components separately. Our numerical analysis reveals that the z-domain (i.e., vertical estimation) can be very sensitive to measurement errors (more than the horizontal estimation) and therefore lead to errors in the z-domain and the overall estimated 3D position. In other words, a small measurement error may lead to a significant positioning estimation error, which is necessarily coming from the estimated position vertical component. As shown in Fig.~\ref{fig_cpacc}-b, we see that high positioning accuracy of $\approx$ 1.89 cm can be achieved in the x-y plane (i.e., horizontal accuracy) for 90\% of UEs. This confirms that the carrier phase method can achieve a high-precision positioning accuracy which is much better than the 30 cm target for horizontal error in the current Rel-17. Fig.~\ref{fig_cpacc}-b also demonstrates that the z-domain (i.e., vertical accuracy) has much larger errors of $\approx$ 13 cm for 90\% of UEs. This reveals that the 3D positioning (based on the phase measurements generated using the absolute time delay model~\cite{38901}) is quite sensitive to the vertical/height estimation error.

\subsection{Positioning Sensitivity to Integer Ambiguity Errors }\label{subsec_IAerr}
Now, we investigate the sensitivity of the horizontal and vertical positioning estimation accuracy to the integer ambiguity errors in the carrier phase measurements. We present a simplified error model for the integer ambiguity problem that follows a discrete uniform distribution on the interval $\left[-{N_{}}_{\text{t}},\,{N_{}}_{\text{t}}\right]$. Here, ${N_{}}_{\text{t}}$ is an integer multiple of the ideal integer ambiguity value (${N_{}}_{\text{e}}$) and is given by ${N_{}}_{\text{t}}=\eta{N_{}}_{\text{e}}$ where $\eta$ is an arbitrary integer with $0<\eta$. Specifically, if a particular link has a wrong integer ambiguity fixing then the value of the error chosen uniformly at random between $\left[-{N_{}}_{\text{t}},\,{N_{}}_{\text{t}}\right]$. Note that ${N_{}}_{\text{e}}$ is defined as the exact integer number of complete phase cycles that the reference signal has travelled between the gNB and UE to produce the observed phase measurement at the UE. The probability of wrong integer ambiguity fixing is then calculated by $\zeta=E/T$ where $E$ is the number of phase measurements with integer ambiguity errors and $T$ is the total number of collected phase measurements.

\begin{figure}
  \begin{center}
  \includegraphics[width=7cm,height=7cm,,keepaspectratio]{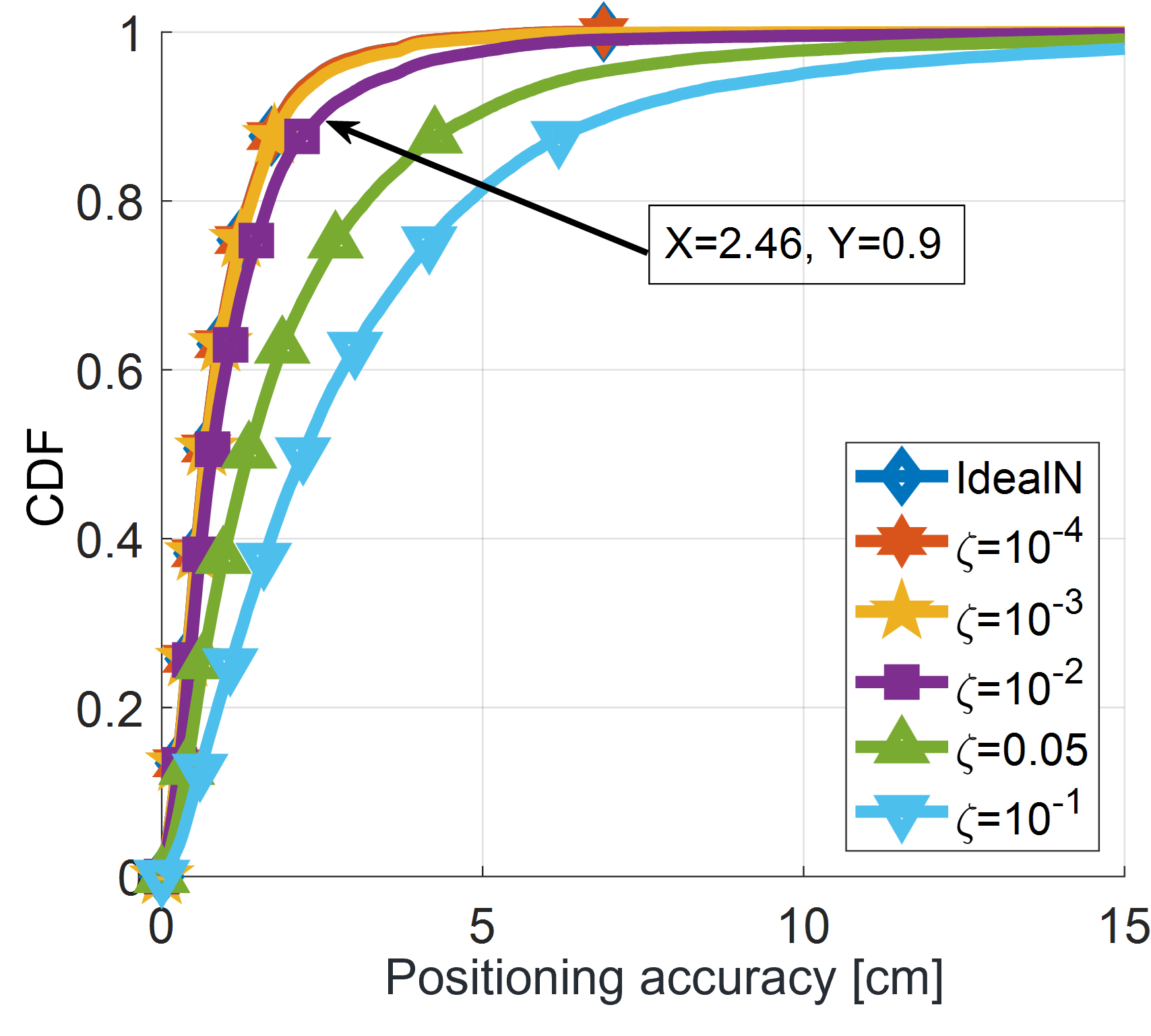}
  \caption{Sensitivity of horizontal accuracy to ambiguity errors.}\label{fig_IA2Derr}
  \vspace{-.25in}
  \end{center}
\end{figure}

\begin{figure}
  \begin{center}
  \includegraphics[width=7cm,height=7cm,,keepaspectratio]{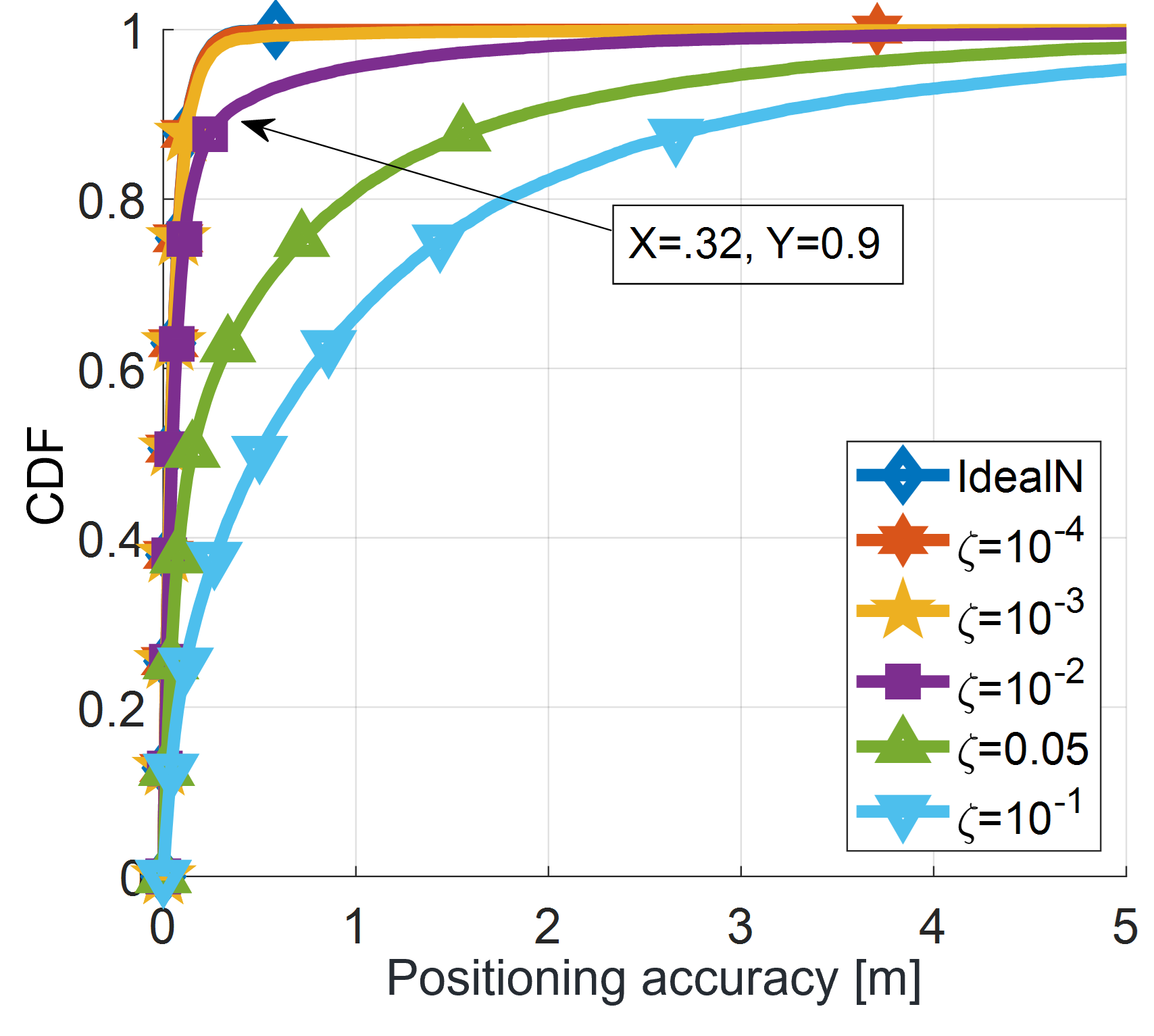}
  \caption{Sensitivity of vertical accuracy to ambiguity errors.}\label{fig_IA3derr}
  \vspace{-.25in}
  \end{center}
\end{figure}

We define $\eta=$3 and $\eta=$23 for the horizontal and vertical positioning estimations, respectively. These values were chosen as reasonable search spaces (i.e., maximum errors) around the true value (i.e., ${N_{}}_{e}$) based on the performance of the timing measurements reported in~\cite{qcom} to restrict the possible search area. As shown in Fig.~\ref{fig_IA2Derr}, the horizontal accuracy degrade as the probability of wrong integer ambiguity fixing ($\zeta$) increases. At $\zeta=10^{-2}$, the horizontal accuracy degrades from 1.89 to 2.46 cm (factor of $\approx$ 1.3) for 90\% of the UEs. The same trend is observed when the sensitivity of the z-domain positioning component (i.e., vertical accuracy) to the integer ambiguity errors is evaluated as shown in Fig.~\ref{fig_IA3derr}. In particular, the vertical accuracy suffers from a substantial deterioration from 13 to 32 cm (factor of $\approx$ 2.4) for 90\% of the UEs when $\zeta=10^{-2}$. This on one hand confirms the sensitivity of the vertical estimation accuracy to the phase measurement errors. On the other hand, this (despite this deterioration) demonstrates the potentiality of the carrier phase method to provide the NR systems with a high-precision (up to cm-level) positioning accuracy compared to the traditional time and angular methods~\cite{Nokia}. 

As discussed in~\cite{Nokia}, the accuracy of positioning measurements (which have been generated using the same simulation settings as discussed in Table~\ref{tab_simpara}) deteriorates substantially when the time measurements are used with the critical and commercial uses cases (e.g., IIoT). In particular, the DL-TDOA method can only achieve a horizontal positioning accuracy of 1.65 m at 90\% of UEs for the InF-SH scenario. In other words, the current RAT-dependent positioning methods may fail to meet the performance requirements defined by the 3GPP Rel-18 for the horizontal and vertical positioning at different service levels~\cite{22261}. Hence, innovative positioning techniques (e.g., carrier phase) need to be considered in the future 3GPP releases to improve the NR positioning accuracy.

\section{Conclusion}\label{sec_conc}
The use of carrier phase-based positioning is investigated in this paper to improve the positioning estimation accuracy in 5G-advanced (Rel-18 and beyond) NR networks. The implementation of the single/double differencing operations and the integration of the reference networks nodes have been discussed to mitigate the impacts of imperfect networks synchronization and enable gNBs and UEs to conduct the required carrier phase measurements. The presented carrier phase positioning technique is evaluated via extensive system-level simulations using the state-of-the-art industrial 3GPP channel model for InF-SH scenarios. Our analysis reveals that the vertical/height positioning estimation has a higher sensitivity to the phase measurements errors (including the integer ambiguity errors) than that of the horizontal estimation. Despite that, the numerical results show that the carrier phase method can still significantly improve the positioning estimation accuracy compared with the current Rel-16/17 methods.

\balance
\bibliographystyle{IEEEtran}
\bibliography{IEEEabrv.bib,Bibliography.bib}
\end{document}